\documentclass[
 longbibliography,
superscriptaddress,
 amsmath,amssymb,
 aps,
 twocolumn,
]{revtex4}

\usepackage{graphicx}
\usepackage{dcolumn}
\usepackage{bm}

\usepackage{natbib}
\begin{document}

\preprint{APS/123-QED}

\title{IceCAPA: patterning particles and microorganisms at a freezing front}

\author{Isabelle M. Feller}
\affiliation{Department of Materials, ETH Z\"{u}rich, Vladimir-Prelog-Weg 1-5/10, 8093, Z\"{u}rich, Switzerland
}
\author{Jakob Paulsen}
\affiliation{Department of Materials, ETH Z\"{u}rich, Vladimir-Prelog-Weg 1-5/10, 8093, Z\"{u}rich, Switzerland
}
\author{Muriel Scherer}
\affiliation{Department of Materials, ETH Z\"{u}rich, Vladimir-Prelog-Weg 1-5/10, 8093, Z\"{u}rich, Switzerland
}
\author{Robert W. Style}
\email{robert.style@mat.ethz.ch}
\affiliation{Department of Materials, ETH Z\"{u}rich, Vladimir-Prelog-Weg 1-5/10, 8093, Z\"{u}rich, Switzerland
}
\author{Lucio Isa}
 \email{lucio.isa@mat.ethz.ch}
\affiliation{Department of Materials, ETH Z\"{u}rich, Vladimir-Prelog-Weg 1-5/10, 8093, Z\"{u}rich, Switzerland
}

\date{\today}

\begin{abstract}
The ability to precisely pattern micro- and nano-scale objects on surfaces is important for a range of different applications. For example, colloidal patterning has been used to create plasmonic surfaces, light-emitting diodes or authentication marks, while microbial cell patterning can be applied to screening antibiotic response and cellular interactions over larger populations at the single-cell level. However, we still lack versatile techniques that can pattern a wide range of synthetic and biological objects on a spectrum of different substrate types. Here, we present a robust patterning technique based on the directional freezing of a particle or bacterial suspension over a patterned substrate. Growing ice pushes the desired objects into traps in the substrate, while sweeping away non-trapped ones, leaving behind high-fidelity patterns. We show that this method works for a range of different materials (both synthetic particles and microbial cells), and is unaffected by substrate wettability. Furthermore, patterned bacterial cells retain excellent post-assembly viability, highlighting the gentle nature of the assembly technique. Beyond patterning applications, our results also give insights into processes involving the freezing of particulate suspensions. In particular, we demonstrate the importance of the temperature gradient as a key control which determines how particles interact with freezing fronts. Finally, we highlight a tight analogy between particles interacting with a freezing front and with air-water interfaces, suggesting that results from capillarity may shed light on freezing phenomena.
\end{abstract}

\maketitle


\section{\label{sec:level1}Introduction}

The precise placement of micron- and nanoscale objects on surfaces plays a key role in a large number of materials and applications. These include the realization of templates for the epitaxial growth of colloidal crystals\cite{vanBlaaderen1997, Lee2004}, for nanoparticle printing\cite{Kraus2007} and for the assembly of colloidal clusters \cite{Ni2016}, or the fabrication of authentication markers \cite{meijs2023}, plasmonic surfaces \cite{Kravets2018}, biosensors \cite{YAP2007775}, and organic light-emitting diodes \cite{LIN2026}.

However, precise particle assembly is challenging. Most self-assembly approaches rely on tuning colloid-scale forces \cite{Hueckel2020}. If these forces are too strong, the particles typically lock together to form disordered structures. If they are too weak, thermal fluctuations prevent structure formation. The delicate nature of colloidal forces makes therefore such processes very sensitive to environmental conditions and particle properties, especially at solid surfaces \cite{Lin2000}. Thus, it is very challenging to make complex patterns via self assembly – especially involving multiple different particle types – and existing protocols are highly system specific.

A more robust approach is to use stronger, longer-ranged external forces for directed assembly \cite{Grzelczak2010}. Among those, capillary forces are at the heart of established particle assembly methods, such as convective assembly \cite{Fleck2015} or Capillarity-Assisted Particle Assembly (CAPA) \cite{Malaquin2007, Ni2018}. The latter is particularly powerful for the deposition of complex, multi-material particle patterns in essentially any 2D configuration \cite{Flauraud2016}. In CAPA, the contact line of an evaporating droplet of a particle suspension is dragged over a topographically-patterned substrate, which contains cavities (or traps) at targeted positions. As the droplet evaporates, evaporative fluxes accumulate particles at the receding air-water meniscus, and some of these particles are pushed into the underlying traps. The moving meniscus then sweeps along particles that are not in traps, leaving only templated, trapped particles behind \cite{Ni2015, Yin2001}. This affords selective depositions with close to 100\% yield without non-specific adsorption over centimeter-squared areas \cite{Kuemin2011}. Upon careful design of the trap arrays, the process can also be repeated sequentially (sCAPA) to fabricate multi-material structures or patterns \cite{Ni2016, Ni2017}.

However, in spite of its success, CAPA suffers from two key limitations. Most importantly, it only works on moderately hydrophilic substrates, with receding contact angles, $30^\circ<\theta_r<60^\circ$, or when surfactants are used to tune the angle to this range \cite{Malaquin2007}. This is because its effectiveness is determined by the direction of the capillary force, which is defined by $\theta_r$. If $\theta_r$ is too large (i.e. approaching $90^\circ$), the capillary force on particles is predominantly parallel to the substrate and this can dislodge particles from traps. If $\theta_r$ is too small (i.e. approaching $0^\circ$), then particles outside traps are pushed strongly down onto the substrate, and friction can prevent them from being swept along with the receding contact line. This leads to non-specific depositions. Thus CAPA can only be performed with intermediate droplet contact angles.

The contact-angle sensitivity of CAPA prevents its application to many materials of high technological interest. In particular, applications such as patterning of cells to study antibiotic susceptibility \cite{Boggon2023}, designing synthetic microbial communities \cite{Johns2016, Bottacin2026} and single-cell responses to antigen stimulation \cite{Yamamura2005} rely on positioning biological objects on highly wetting substrates beyond the applicability of CAPA. For such applications, cell surface attachment is often mediated by coatings that promote adhesion \cite{Seo2025, ZHANG20094021}, and surface passivation and prevention of non-specific binding are achieved by anti-fouling coatings \cite{Asuri2007}. These coatings are often hydrophilic \cite{Banerjee2011}, inducing complete wetting ($\theta_r=0$), and thus preventing the use of CAPA for cell patterning. This is also true of hydrogels \cite{Kim2021} -- another type of substrates commonly used in cell culture \cite{Kandemir2018}.

CAPA’s second key limitation is desiccation, which greatly hinders its use for biological objects. Drying is an integral part of the process and after the receding meniscus moves, objects deposited in the traps dry in ambient conditions. While desiccation poses little problems for some biological objects, it can be fatal for many others, including common bacteria lab strains \cite{Pioli2021,potts1994desiccation}.

Here we propose an alternative to conventional CAPA that overcomes these limitations by exploiting a completely different system, but which largely shares the same physics. This approach, IceCAPA, exchanges the water-interface of a receding droplet with an advancing ice/water solidification front. The growing ice pushes particles ahead of it \cite{KORBER1985649}, allowing us to achieve similar particle templating results to CAPA, while avoiding drying and exposure to air.

\section{\label{sec:level2}Results and discussion}

\begin{figure}[h!]
    \centering
    \includegraphics[width=1\linewidth]{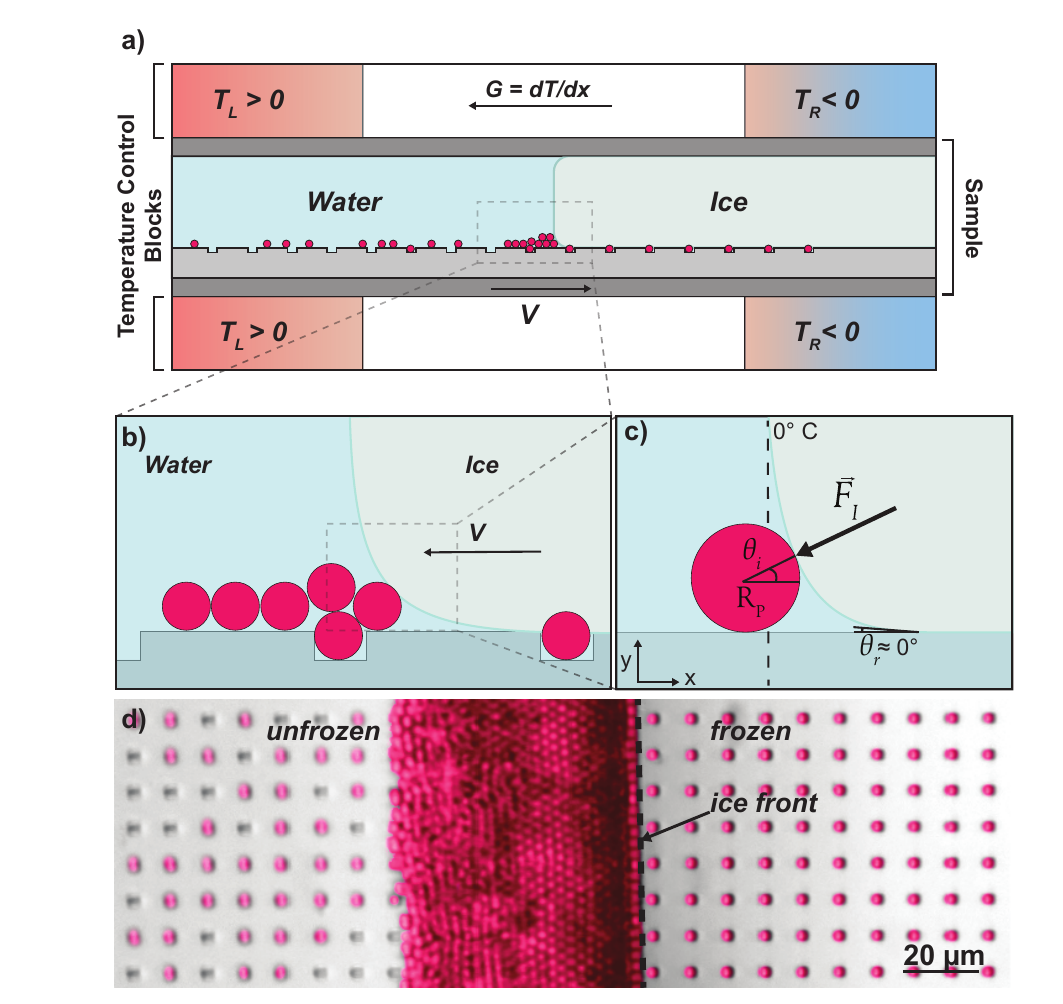}
    \caption{The working principle of IceCAPA. a) Schematic of the experimental setup. b,c) Schematics showing how ice pushes particles ahead of it, leading to particles being deposited in traps, and illustrating the angle, $\theta_i$ of the force $\overrightarrow{F_I}$ exerted by an ice front on a single particle. d) Micrograph of an IceCAPA experiment (20x air objective, Nikon Eclipse Microscope), showing ice growing through a suspension of silica particles in ultrapure (18.2 M$\Omega$.cm) water ($R_p=1.5\,\mu$m), pushing a dense packing of particles.}
    \label{Figure1}
\end{figure}

To perform IceCAPA, we freeze a suspension-filled cell using the apparatus shown schematically in Figure \ref{Figure1}a (see \cite{gerber2022stress,gerber_stage} for details). This apparatus allows us to grow ice through the cell, while independently controlling the temperature gradient, $G$, and the freezing speed, $V$. The sample cell consists of two glass slides, separated by spacers, with a substrate (patterned with traps) placed on the bottom glass slide. The cell's right end is sandwiched between a cold pair of copper blocks which are set to a temperature $T_l<0^\circ$C, while the left end is sandwiched between a warmer pair of copper blocks with $T_r>0^\circ$C. This results in a freezing front forming in the cell where the local temperature within the $w=2.4$ mm gap between the blocks reaches $T=0^\circ$C, and which can be visualized with an optical microscope. Fixing the block temperatures also sets up a fixed temperature gradient of $G=(T_r-T_l)/w$ on the portion of the sample around the freezing front. To move an ice front through a liquid-filled sample cell, we push the cell towards the subzero block at a constant speed, $V$. This causes an ice front to grow steadily through the suspension at the same speed. In the frame of reference used for imaging the sample, the ice/water interface appears stationary (at the  0$^\circ$C isotherm).

As ice moves through the suspension, it readily pushes particles ahead of it, allowing us to deposit them into the traps \ref{Figure1}b-c. In Figure \ref{Figure1}d we show how 1.5 $\mu$m-radius, silica particles pile up in front of a growing ice front, leading to a dense accumulation zone (here, $G=2\,$K/mm, $V=40\,\mu$m/min), analogous to the accumulation observed in CAPA. Here, the substrate is PEGylated silicone (see Methods), with 3.2 $\mu$m wide, 3.7 $\mu$m long and 1.5 $\mu$m deep traps. As the accumulation zone slides over the substrate, it pushes one particle into each unfilled trap, leaving an almost perfect particle array on the frozen side of the cell. With this set of freezing parameters, there are no particles deposited outside of the pattern, highlighting the precision and fidelity of this method. As in conventional CAPA, IceCAPA also allows for the sequential deposition of multiple particles in the same trap, i.e. by replacing the particle suspension and moving a second ice front over the template (see SI - Figure S1)).

\subsection{Parameter optimization in IceCAPA}
While Figure \ref{Figure1}d demonstrates how near perfect depositions can be achieved, there are a number of parameters that must be optimized to systematically achieve good particle assembly. Firstly, we need to use freezing speeds that are lower than the critical freezing speed, $V_c$, above which particles are engulfed by growing ice (in the absence of traps). $V_c$ depends on the particle type, size, and the substrate properties. For example, it is well established that $V_c$ is much smaller for larger particles than smaller ones \cite{REMPEL2001420, LIPP1993475}. Here, for a PEGylated silicone substrate, 1.4 $\mu$m radius polystyrene particles are engulfed for $V>\,$100 $\mu$m/min, and silica particles with a radius of 1.5 $\mu$m are engulfed for $V>\,$280 $\mu$m/min. Qualitatively, we found that  lower friction substrates had higher $V_c$. To avoid unwanted particle engulfment at the freezing front, in all experiments, we kept $V<\,$60 $\mu$m/min.

\begin{figure*}[h]
    \centering
    \includegraphics[width=0.7\linewidth]{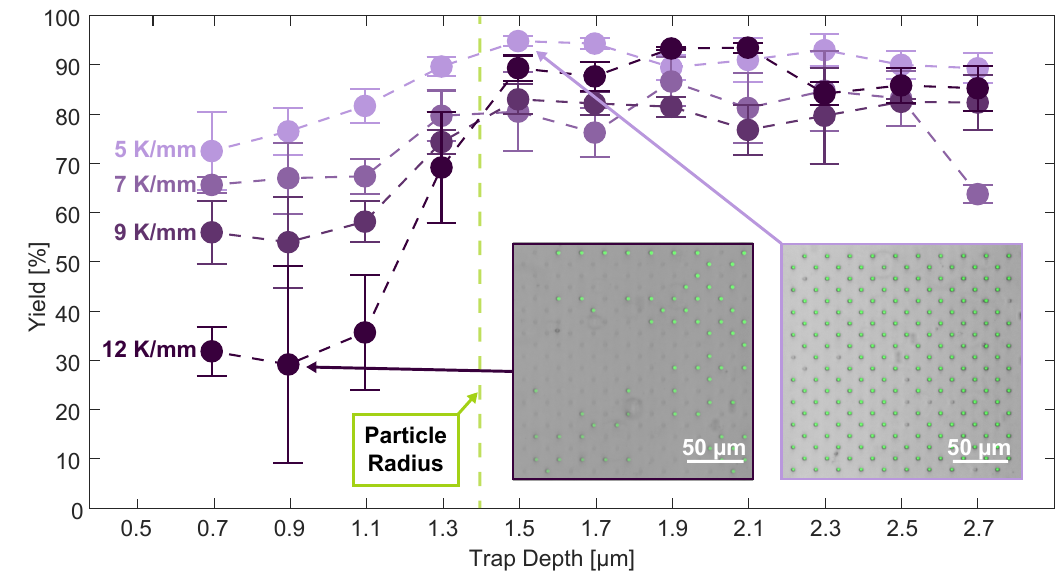}
    \caption{IceCAPA success depends on temperature gradient and trap size. The plot shows yield (\% of correctly placed particles) as a function of trap depth. The different color datasets show the results for different temperature gradients, $G$. All experiments use polystyrene particles with $R_p=1.4 \,\mu$m on silicone substrates. We repeated each experiment at least three times -- inter-experiment variability is shown by the size of error bars. Insets show brightfield images of deposited particles for two different sets of conditions.    }
    \label{Figure2}
\end{figure*}

Secondly, the trap depth also strongly affects IceCAPA's success. For example, Figure \ref{Figure2} shows the yield (\% of correctly placed particles), $Y$, for polystyrene particles with a particle radius $R_p=1.4\,\mu$m being pushed into traps of different depths, $d$. The results are shown for a range of different $G$. In all cases, $Y$ depends strongly on the depth of the traps relative to $R_p$. For $d>R_p$, $Y$ is consistently high. For smaller $d$, $Y$ drops off significantly. This is to be expected, as particles can be more easily dislodged from traps when a particle's waist sticks out above the top of the trap \cite{Ni2015}. Thus, successful IceCAPA requires $d>R_p$ -- similar to the requirement for standard CAPA that $d \approx R_p$ \cite{Ni2016,Flauraud2016}). We do not vary trap lateral dimensions here: we always use trap widths $\approx 2.2 \, R_p$, following our experience that these produce good results for standard CAPA.

Finally, the applied temperature gradient can strongly affect yield. Figure \ref{Figure2} shows the effect on $Y$ of varying $G$ between 5 and 12 K/mm. For shallow traps ($d<R_p$), yield is low for high temperature gradients $(G=12$ K/mm) with many traps remaining unfilled (Figure \ref{Figure2}). However, $Y$ increases significantly as $G$ decreases to $5$ K/mm.

\onecolumngrid

\twocolumngrid

This dependence upon temperature gradient is a consequence of the local shape of the ice-water meniscus at the contact line, and how this changes how particles are pushed by growing ice (see schematic in Figure \ref{Figure1}c). This meniscus shape actually satisfies the same equation as the equation for the shape of gravitational liquid menisci at substrates, which they completely wet. However, whereas gravitational menisci are controlled by a balance between surface tension and gravity, solid-liquid menisci are controlled by a competition between the solid-liquid interfacial energy and ice’s drive to freeze when undercooled (the Gibbs-Thomson equation). This competition defines a characteristic \emph{thermocapillary} length scale, $L$, over which the ice-water meniscus is rounded out at the contact line \cite{wilen1995giant}: $L=\sqrt{\gamma T_m/(\rho q_m G)}$,
where $G = \frac{dT}{dx}$ is the temperature gradient, $T_m$ is the ice melting temperature, $q_m$ is the specific latent heat of melting, $\rho$ is the liquid density, and $\gamma$ is the interfacial energy of the solid-liquid interface. In particular, the 2D-shape of the meniscus near the contact line is well approximated by $y=L e^{-x/L}$ (see Methods \ref{meniscus shape} for the full expression) \cite{wilen1995giant}. Thus, large $G$ will lead to small menisci, and vice versa. Note the strong analogy between $L$ and the capillary length $\lambda=\sqrt{\gamma/(\rho g)}$, which describes meniscus shapes at air-water interfaces (where $\gamma$ is water's surface tension, and $g$ is the gravitational acceleration). 

\begin{figure}[h!]
    \centering
    \includegraphics[width=1\linewidth]{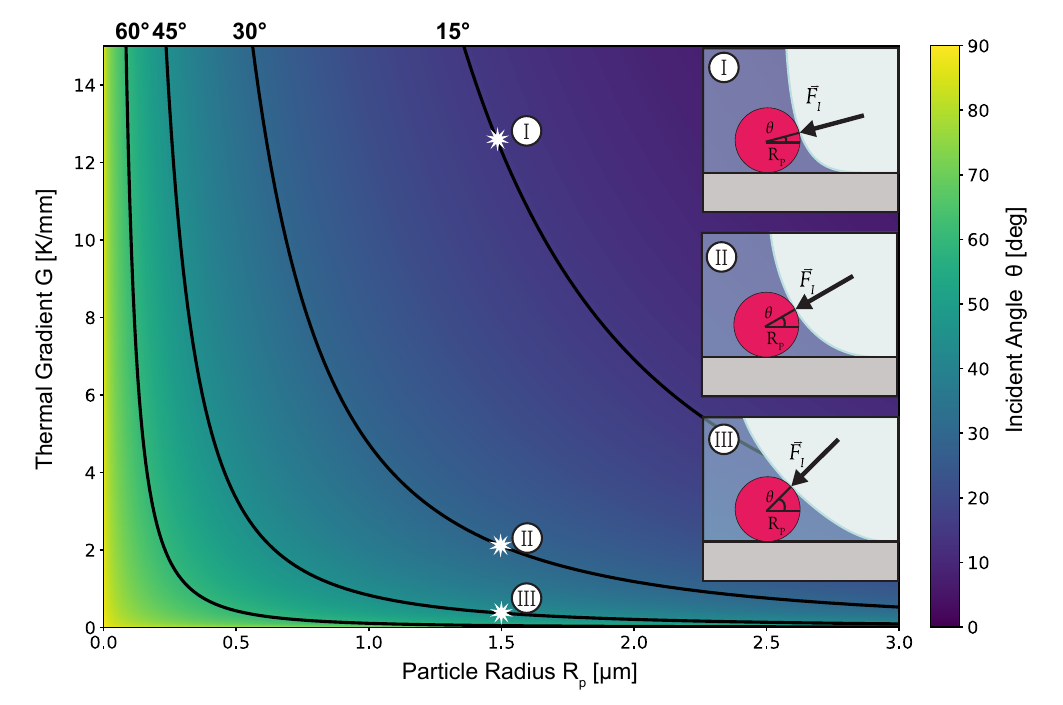}
    \caption{Contour plot showing how incidence angle, $\theta_i$, depends on particle radius $R_p$ and the applied thermal gradient $G$. Solid black curves indicate isolines of constant $\theta_i$. The white star markers correspond to the three specific scenarios shown in the insets on the right.}
\label{Figure3}
\end{figure}

The relative size of $L$ to the particle radius, $R_p$ dictates how growing ice will push a particle. When $L\ll R_p$ (larger $G$), particles are much larger than the thermocapillary meniscus, and so essentially just feel a planar ice front advancing laterally across the substrate (e.g. Figure \ref{Figure3} inset I). Assuming that the ice exerts a force on the particle orthogonal to its surface, this means that particles are pushed more tangentially across the substrate. This lateral force can easily push particles out of shallow traps, leading to many unfilled traps (e.g. results for $G=12$ K/mm in Figure \ref{Figure2}). By contrast, when $L>R_p$ (smaller $G$), particles feel the curvature of the thermocapillary meniscus (Figure \ref{Figure3} inset III). Now the ice meets the particle at an angle that pushes particles down into the substrate, enabling efficient trapping (e.g. results for $G=5$ K/mm in Figure \ref{Figure2}).

In fact, we can use calculated meniscus shapes to predict suitable values of $G$ for performing IceCAPA. In the Methods, we calculate the incidence angle, $\theta_i$, that ice pushes particles with, by solving the Gibbs-Thomson equation and assuming that water completely wets ice substrate interfaces (a good approximation for most materials \cite{sarlin2025macroscopic,murata2016thermodynamic,slater2019surface,baran2025understanding}). The results are shown in Figure \ref{Figure3}, and show how $G$ strongly affects $\theta_i$. Indeed, for $R_p=1.5\,\mu$m, $\theta_i$ changes by several tens of degrees as $G$ increases from 1 to 10$\,$K/mm. This explains the large differences in experimental behavior as we tune $G$ in this range (Figure \ref{Figure2}). More generally, Figure \ref{Figure3} shows that tuning $G$ is an easy way to tune the effective contact angle of the ice-water meniscus that pushes particles. When speed, trap size, and temperature gradient are optimized, we obtain the almost perfect patterning shown in Figure \ref{Figure1}d (see SI - MovieS1).

\subsection*{Wettability-independent deposition}

\begin{figure}[h!]
    \centering
    \includegraphics[width=1\linewidth]{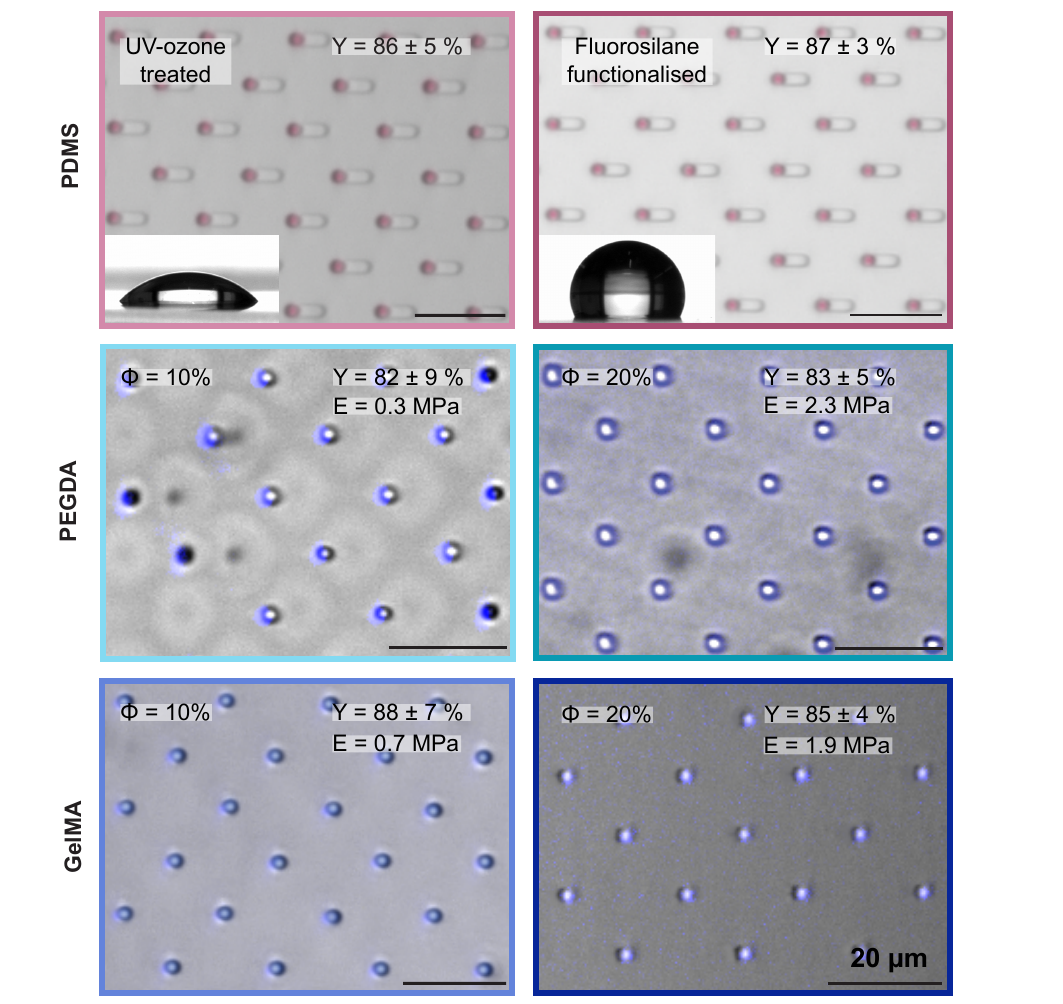}
    \caption{IceCAPA deposition of silica particles with $R_p= 1.5 \,\mu$m (top row) and $R_p= 1 \,\mu$m (middle and bottom row). $G=1\,$K/mm and $V=$50 $\mu$m/min. Top row: Hydrophilic, UV-ozone treated silicone (left) and hydrophobic, fluorosilane-functionalized silicone (right).
    The insets show water droplets on the functionalized surfaces, demonstrating the difference in wettability.
    Middle row: Polyethylene Glycol Diacrylate (PEGDA) hydrogels with polymer contents of $10\,\%$ (w/v) (left) and $20\,\%$ (w/v) (right). Bottom row: Gelatin methacryloyl (GelMA) hydrogels with polymer contents of $10\,\%$ (w/v) (left) and $20\,\%$ (w/v) (right). Water completely wets all the hydrogel substrates ($\theta_r=0$). Reported Young's moduli, $E$, are measured using atomic force microscopy.}\label{Figure4}
\end{figure}

Importantly, unlike conventional CAPA \cite{Malaquin2007, Ni2015}, IceCAPA works independently of the substrate wettability. This allows us to pattern particles on a wide range of different substrates that cannot be patterned by standard CAPA. We demonstrate this substrate agnosticism by performing particle assembly on a range of different surfaces. The two top panels in Figure \ref{Figure4} show IceCAPA results on functionalized silicone substrates that are surface treated with UV-ozone/fluorosilane to make them hydrophilic/hydrophobic, respectively \cite{Glass2011} (insets show water droplets on the surfaces with respective contact angles of $44\,\pm \,2 \,^\circ$ and $105 \,\pm \, 2\, ^\circ$). In both cases, we achieve excellent patterning, despite the huge difference in substrate wettability.

IceCAPA even works on completely wetting substrates like hydrogels (see SI - MovieS2). The four bottom panels in Figure \ref{Figure4} show IceCAPA results using hydrogels made from poly(ethylene glycol) diacrylate (PEGDA) and gelatin methacrylol (GelMA) using polymer volume fractions of $\phi=10$ or $20\,\%$. For all these surfaces, we again achieve good particle assembly. Indeed, the process works, despite the fact that the hydrogels are soft (with $O$(MPa) Young's moduli, $E$), so that stresses from the growing ice front can potentially deform them to push particles out of traps \cite{gerber2023polycrystallinity,feng2025characterizing} (see SI - Figure S2 for additional details).

\newpage 

\subsection*{Assembling bacterial cells on hydrogels}
\begin{figure}[h!]
    \centering
    \includegraphics[width=1\linewidth]{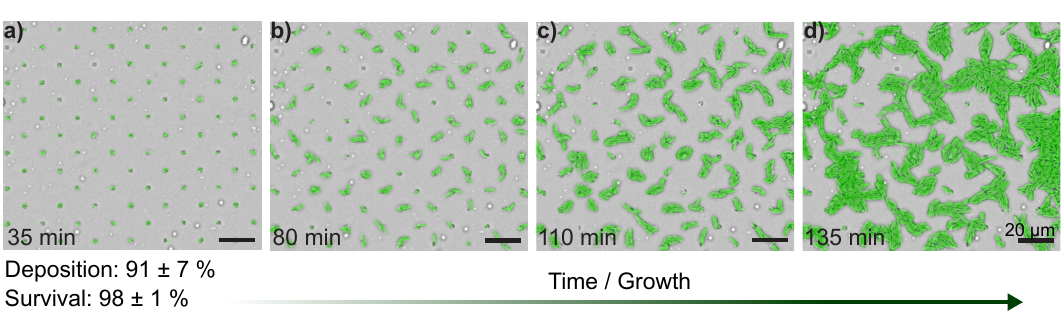}
    \caption{Time-lapse images of \textit{E. coli} MG1655 cells deposited on a GelMA 10$\,\%$ hydrogel. The cells were deposited using IceCAPA with $G=$1.5 K/mm and $V=$30 $\mu$m/min. After the deposition we melted the ice, transferred the gel to a glass-bottom petri dish, placed 1.5$\,\%$ Luria Bertani agar (LBA) on top of the deposited cells and held them at 37 $^\circ$C in a heated microscopy chamber. The images show stages of growth during this incubation step ($t=0$ at the point of LBA addition).
    }\label{Figure5}
\end{figure}

The successful patterning of particles on hydrogels opens up new possibilities for the patterning and study of biological objects, such as bacterial cells. We showcase this potential in Figure \ref{Figure5}, by demonstrating precision patterning of \textit{Escherichia coli} bacteria (1 by 2 $\mu$m rod-shaped cells) on a GelMA hydrogel. Unlike in conventional CAPA (where cells are dried out when exposed to air, resulting in a survival rate of $0\, - \,45 \,\%$ after patterning \cite{BoggonPhD2024, Pioli2021}), we can maintain high cell viability. This is because IceCAPA keeps the cells hydrated throughout the patterning process. Although there are other methods providing high cell viability, i.e. droplet-based approaches \cite{Xiong2025} or single-cell isolation via centrifugation \cite{Whitley2022}, combining a high-throughput method with single-cell resolution still remains challenging, and IceCAPA offers a robust solution.
To prepare the cell suspension for deposition, we grow \textit{E. coli} cells overnight in liquid media, and then dilute them into $10\,\%$  phosphate-buffered saline (PBS) solution. We then deposit the cells in $2\times 2\times 2$ $\mu$m$^3$ traps via IceCAPA with $V=30\, \mu$m/mm and $G$ = 1 K/mm. The resulting yield is $91\,\%$, with no bacteria left on the hydrogel surface outside of the traps (Figure \ref{Figure5}a). After the deposition, we melt the ice, exchange the cell suspension with growth media and incubate the sample at 37 $^\circ$C in a temperature-controlled microscopy chamber. The cells maintain a high viability of 98\%, as demonstrated by the subsequent cell growth (Figure \ref{Figure5}b-d and SI - MovieS3).

\section{Conclusions}

Taken altogether, our results show that IceCAPA is a powerful tool for precise micropatterning of synthetic particles and microorganisms on surfaces, and that it overcomes the limitations of previous techniques. In particular, IceCAPA allows patterning of microscale particles on any substrate -- provided that they do not stick to each other or to the substrate. It presents substantial advantages for use with biological materials, as it does not involve desiccation caused by air exposure and mild environmental conditions. Furthermore, IceCAPA can be tuned \emph{in situ}, by simply changing $G$ and $V$, to quickly optimize deposition parameters. In principle, IceCAPA can also be performed with any other solvent that easily rejects particles while freezing (e.g. camphene or tert-Butyl alcohol \cite{deville2008freeze}). Thus, it should be straightforward to adapt to hydrophobic particle systems by using more hydrophobic solvents. Finally, upon thawing the ice, particle and cell patterns can be exposed to different liquid environments for further downstream processes (see SI - Figure S3), such as particle binding or bacterial growth assays.

Beyond IceCAPA, our results give insights into the behavior of freezing particle suspensions (e.g. as found in cell freezing \cite{lovelock1953haemolysis}, frost heave \cite{teng2023frost,peppin2013physics}, food technology \cite{deville2017freezing}, freeze structuring \cite{deville2018lure,ruhs2026freeze}, and water treatment \cite{vesilind1990freezing}). For example, we have shown how the shape of a growing ice-water interface can control whether particles are engulfed during freezing or not -- with particles being easily engulfed near large ice-water menisci. The size, $L$, of these menisci is set by the temperature gradient at the freezing front, with $L\sim G^{-1/2}$. Thus, particle engulfment by ice occurs more easily in freezing conditions with low temperature gradients. Furthermore, our results highlight the strong analogy between freezing and problems involving dense particles at fluid interfaces: in both cases the interface shapes satisfies the same meniscus equation. The only difference is that the gravitational acceleration $g$ is replaced by an effective gravitational constant, $g'=q_m G/T_m$ for the case of an ice-water interface. This analogy implies that we may be able to gain further insights into freezing suspensions by applying results from capillary phenomena like particle sinking \cite{vella2015floating,protiere2017sinking}, the Cheerios effect \cite{meijer2025frozen, vella2005cheerios} and dip coating \cite{gans2019dip}. An interesting question for future work will be to determine how far the analogy holds, and how it is affected by complexities such as faceting of solid-liquid interfaces \cite{wettlaufer1999crystal}, and morphological instabilities that can appear during solution freezing \cite{mullins1964stability,peppin2007morphological,you2016interfacial}.

\section{Methods}

\subsection{Incidence angle of ice pushing on a particle}\label{meniscus shape}
To calculate the incidence angle of an ice front pushing a particle, we use the fact that the shape of an ice-water meniscus is given by the same equation that describes a fluid-fluid gravitational meniscus \cite{wilen1995giant} -- but with gravity replaced by $q_m G/T_m$, where $q_m=3.34\times10^5$ J/kg is the specific latent heat of freezing of water, $G$ is the temperature gradient, and $T_m=273$ K is the bulk freezing temperature of water \cite{hobbs2010ice}.
\emph{N.B.}, this assumes that ice has an isotropic surface energy.
Furthermore, ice has an effective contact angle of $\approx 0^\circ$ against most materials (as shown in Figure \ref{Figure1}c) \cite{sarlin2025macroscopic,murata2016thermodynamic,slater2019surface,baran2025understanding}.
Thus, the shape of the meniscus is given by \cite{landau1987fluid}:
\begin{equation}
    \frac{x}{L}=\cosh^{-1}\left(\frac{2L}{y}\right)-2\sqrt{1-\left(\frac{y}{2L}\right)^2}-\cosh^{-1}\sqrt{2}+\sqrt{2}.
    \label{eqn:meniscus}
\end{equation}
Here, $L=\sqrt{\gamma T_m/\rho q_m G}$, $\gamma=0.033\, \mathrm{J}/\mathrm{m}^2$ is the surface energy of an ice-water interface \cite{hobbs2010ice}, $x=0$ corresponds to the $0^\circ$C isotherm, and $y=0$ is the substrate surface (see Figure \ref{Figure1}c).
This equation is well approximated by $y/L=e^{-x/L}$ for $y<0.6L$ (i.e. near the substrate surface).

We assume that the ice pushes on the particle at an angle orthogonal to the ice/water interface. Thus, the pushing incidence angle is related to the meniscus slope by:
\begin{equation}
    -\frac{dx}{dy}=\cot\theta.
\end{equation}
Inserting this into equation (\ref{eqn:meniscus}) gives $\theta(y/L)$.
Secondly, the geometry of the particle touching the meniscus requires that
\begin{equation}
    \frac{y}{L}=\frac{R_p}{L}(1+\sin\theta).
\end{equation}
This allows us to calculate $R_p/L$ as a function of $y/L$.
Eliminating $y/L$ gives $\theta(R_p/L)=\theta(R_p/\sqrt{G})$.
Thus, given $R_p$ and $G$, we can calculate $\theta$, as shown in Figure \ref{Figure3}.

\subsection{Substrate fabrication}\label{traps}
For substrate fabrication, we use all chemicals as received.
We prepare molds to make substrates with traps, by printing cuboid shapes on fused-silica slides.
Briefly, we design arrays of shapes to be printed using Tinkercad and process them with  DeScribe (NanoScribe GmbH, Germany). 
We then prepare the silica slides (Multi-Dill, NanoScribe GmbH, Germany) by cleaning them with ethanol and plasma-treating them for $20\, \mathrm{s}$ with a Piezobrush PZ2 (relyon plasma GmbH).
Finally, we print the shapes by using two-photon polymerization (Photonic Professional GT2, NanoScribe GmbH, using a Nikon 63x, NA 1.5 objective) to locally crosslink IP-Dip2 resin (Nanoscribe GmbH).
To remove uncrosslinked resin, we incubate the samples for 15 minutes in PGMEA (Propylene glycol methyl ether acetate, $\geq99.5\,\%$, Sial), followed by 3 minutes in isopropanol.
After this development step, we modify the printed surfaces by using chemical vapor deposition under vacuum for 2h to silanize them with (Trichloro(1H,1H,2H,2H-perfluorooctyl) silane ($97\,\%$, Sigma Aldrich) -- which reduces adhesion during the molding process.
Finally, we bake samples on a hotplate with a temperature that ramps up to 175 $^\circ$C over 2 hours.
This final step helps improve adhesion between the printed resin and the glass slide.

We make poly(dimethylsiloxane) (PDMS) substrates that replicate the printed masters using the silicone elastomer Sylgard 184 (Dow Chemicals).
We prepare a mixture of base and curing agent with a 10:1 ratio, and pour the mixture on top of the printed molds.
After curing overnight at 80 $^\circ$C, we remove the silicone from the mold, and place it, trap-side up, on a glass slide.
To PEGylate the PDMS substrates with polyethylene glycol (PEG), we first UV-ozone clean the substrates for 10 min before pipetting a 1\% (w/v) 3-[Methoxy(polyethyleneoxy)propyl]trimethoxysilane ($90\,\%$, 6-9 PEG units, ABCR) in ethanol solution onto the sample. We let the mixture react on the surface for 30min then we first rinse the samples with ethanol, followed by a rinse with ultrapure water (18.2 M$\Omega$.cm).

We similarly make hydrogel substrates using either poly(ethylene glycol) diacrylate (PEGDA) or gelatin methacryloyl (GelMA) to replicated the printed masters.
These substrates need to be bonded to a glass slide so that they do not move during the IceCAPA process.
Thus, we start by preparing glass slides that are functionalized with methacrylate groups that will bond to the hydrogels.
We plasma treat glass slides for 10 minutes, before immersing them in a mixture of 50 $\mu$L glacial acetic acid ($> 99.8\,\%$, VWR Chemicals), 42 $\mu$L ultrapure water (18.2 M$\Omega$.cm), 905 $\mu$L ethanol and 3 $\mu$L 3-(Trimethoxysilyl)propyl methacrylate(440159, Sigma Aldrich) that has been allowed to rest for 5 minutes after mixing.
We then wash the slides in ethanol, and heat them on a hotplate at 110 $^\circ$C for 1 hour to complete the silanization reaction.

To make PEGDA substrates, we mix 10 or 20 vol$\%$ PEGDA (molecular weight 700 Da, Sigma-Aldrich) with 0.0005$\,$vol$\%$ 2-hydroxy-2-methylpropiophenone ultraviolet initiator in ultrapure water (18.2 M$\Omega$.cm).
We pipette the mixture into a sandwich consisting of the printed mold (on the bottom, printed side up) and the methacrylate-functionalized glass slide (on the top, face down), separated with 170 $\mu$m thick glass spacers.
We then crosslink the gel with 356 nm ultraviolet light (Analytik Jena US) for 1 hour.
After this step, the printed mold easily detaches, leaving a hydrogel substrate supported on a glass slide.

To make GelMA (160kDa, $80\,\%$ degree of modification) substrates, we use the same protocol as for PEGDA substrates.
However, we use 10 or 20 $\,$\%(w/v) (X-pure GelMA 205805RG, BIO INX, Belgium) with respectively 0.1 or 0.2 $\,\%$(w/v) LAP initiator (Lithium phenyl-2,4,6-trimethylbenzoylphosphinate, 85073-19-4, Sigma Aldrich) in ultrapure water (18.2 M$\Omega$.cm).
The mixture is prepared at 35 $^\circ$C in a water bath to fully dissolve GelMA.
We could store the complete un-polymerized mixture at 4 $^\circ$C for up to a month.
Such cold-stored samples solidify into a physical gel, and thus must be warmed back to 35 $^\circ$C to melt them prior to use.

\subsection{Particle and bacterial cell suspensions}\label{Particle_suspension}
For the particle depositions, we use green fluorescent polystyrene particles (1.4 $\mu$m radius, PSFluoGreen- Fi135, Microparticles GmbH) or  silica particles (SiO2) (1.5 $\mu$m radius SiO2-R-LSC84; blue fluorescent 1 $\mu$m radius SiO2-FluoBlue-SC2013, Microparticles GmbH). Before  deposition we wash and disperse the particles in de-ionized water (ultrapure, 18.2 M$\Omega$.cm) to reach a final concentration of 0.1 mg/mL. We chose this concentration since a lower particle concentration leads to a smaller accumulation zone and incomplete filling of the traps, while a higher particle concentration results in a large accumulation zone which cannot be properly pushed by the ice meniscus.
For bacteria depositions, we use cell suspensions of  \textit{E. coli} MG1655 cells (from M. Seeger lab, \cite{Sorgenfrei2025}).
These are plated on a 1.5$\,\%$ Luria-Bertani agar plate and grown overnight at 37 $^\circ$C. 
The next day, we pick single colonies, incubate them in M9 medium (63011-500g-F, Millipore, Sigma Aldrich) containing 0.2$\,\%$(w/v) glucose, 5 mM MgSO4 and 0.1 mM CaCl2 and culture them overnight at 37 $^\circ$C and 200 rpm in a shaker incubator.
The next day, we resuspend the cells in Phosphate-Buffered Saline (PBS), wash the suspension three times, and then dilute it five times in 0.1$\times$ PBS to prepare a dilute bacteria suspension for deposition. 

\subsection{IceCAPA procedure}\label{deposition}
The sample cell consists of a glass-slide-supported silicone or hydrogel substrate on the bottom, and a glass slide on the top, separated by glass or SecureSeal spacers (360 - 600 $\mu$m in height).
We inject a colloidal or bacterial suspension into the sample cell and insert it into the temperature stage. We set the temperature gradient in the temperature stage and nucleate ice in the sample on the subzero side using a cotton swab dipped into liquid nitrogen \cite{gerber2022stress}. Once the ice front is stable in the field of view, we start with the deposition by moving the stage.
For colloidal suspensions, we perform IceCAPA by moving the stage with a speed of 50 $\mu$m/min and a temperature gradient between 1 and 12 K/mm.
For bacterial cells, we use a stage speed of 30 $\mu$m/min and a temperature gradient of 1.5 K/mm.
After bacterial deposition, we immediately defrost the sample and transfer the gel to a glass bottom petri dish where we add 1.5$\,\%$ (w/v) Tryptic Soy Agar (TSA, Millipore). We then transfer the sample into a microscopy chamber heated to 37 $^\circ$C where we record a brightfield image every 5 min for 15 hours using a Nikon Eclipse Ti2 microscope and a 60$\times$ air objective.

\subsection*{Acknowledgements}

We gratefully acknowledge Eleonora Secchi for her advice regarding the culturing conditions to observe the bacterial growth, Yanxia Feng for her support in the fabrication of the hydrogels, Shivaprakash Ramakrishna and Simon Scherrer for their help the AFM measurements, and Nick Stauffer and Xueting Shen for helping with printing the master molds using the Nanoscribe. We furthemore acknowledge the laboratory of Markus Seeger sharing the \textit{E. coli} MG1655 cells. 
IF, MS and RWS acknowledge support from the
Swiss National Science Foundation (200021-212066). This work was carried within the framework of the ALIVE initiative (Advanced Engineering with Living Materials) and funded by the SFA-AM program (Strategic Focus Area – Advanced Manufacturing) by the ETH Board.

\subsection*{Supporting Information}

Further experimental details, including supplemental figures and movies are given in the Supporting Information.

\bibliography{sn-bibliography}

\end{document}